\documentclass[journal]{IEEEtran}
\usepackage{commath,graphicx,afterpage,subfigure,amsmath,amsfonts,amssymb,algorithmic}
\usepackage{mathrsfs,hyperref,amsthm,cite,multirow,bm,multicol,booktabs,threeparttable,extpfeil}
\usepackage{algorithm}
\usepackage{algorithmic}

\begin{document}

\title{RAICL: Retrieval-Augmented In-Context Learning for Vision-Language-Model Based EEG Seizure Detection}

\author{Siyang~Li, Zhuoya~Wang, Xiyan~Gui, Xiaoqing~Chen, Ziwei~Wang, Yaozhi~Wen, and Dongrui~Wu, \IEEEmembership{Fellow,~IEEE}
\thanks{S.~Li, Z.~Wang, X.~Gui, X.~Chen, Z.~Wang, and D.~Wu are with the Ministry of Education Key Laboratory of Image Processing and Intelligent Control, School of Artificial Intelligence and Automation, Huazhong University of Science and Technology, Wuhan 430074, China.}
\thanks{Y.~Wen is with the State Key Laboratory of Brain Cognition and Brain-inspired Intelligence Technology, Institute of Automation, Chinese Academy of Sciences, Beijing 100190, China.}
\thanks{S. Li, Z. Wang and X. Gui contributed equally to this work.}
\thanks{This research was supported by National Natural Science Foundation of China (62525305).}
\thanks{Corresponding Author: D. Wu (drwu09@gmail.com).}}

\markboth{}
{Shell \MakeLowercase{Li~\emph{et al.}}: }
\maketitle

\begin{abstract}
Electroencephalogram (EEG) decoding is a critical component of medical diagnostics, rehabilitation engineering, and brain-computer interfaces. However, contemporary decoding methodologies remain heavily dependent on task-specific datasets to train specialized neural network architectures. Consequently, limited data availability impedes the development of generalizable large brain decoding models. In this work, we propose a paradigm shift from conventional signal-based decoding by leveraging large-scale vision-language models (VLMs) to analyze EEG waveform plots. By converting multivariate EEG signals into stacked waveform images and integrating neuroscience domain expertise into textual prompts, we demonstrate that foundational VLMs can effectively differentiate between different patterns in the human brain. To address the inherent non-stationarity of EEG signals, we introduce a Retrieval-Augmented In-Context Learning (RAICL) approach, which dynamically selects the most representative and relevant few-shot examples to condition the autoregressive outputs of the VLM. Experiments on EEG-based seizure detection indicate that state-of-the-art VLMs under RAICL achieved better or comparable performance with traditional time series based approaches. These findings suggest a new direction in physiological signal processing that effectively bridges the modalities of vision, language, and neural activities. Furthermore, the utilization of off-the-shelf VLMs, without the need for retraining or downstream architecture construction, offers a readily deployable solution for clinical applications.
\end{abstract}


\section{Introduction}

\IEEEPARstart{E}{lectroencephalography} (EEG) analysis is a fundamental tool in neuroscience, enabling the measurement of underlying neural activity via multiple scalp electrodes. Characterized as multivariate time-series data, EEG signals are typically acquired using non-invasive headsets, offering significant advantages in terms of accessibility, cost-effectiveness, and ease of deployment~\cite{Wu2023}. Consequently, the accurate decoding of EEG signals is critical for a wide range of applications, including medical diagnostics, rehabilitation engineering, and brain-computer interfaces~\cite{Gao2025, Edelman2025}.

Conventional signal-based approaches, however, struggle to effectively address the inherent complexities of EEG. These methods often require extensive subject-specific calibration or fail to generalize across diverse sessions, subjects, or tasks. Currently, practical deep learning architectures for EEG decoding remain dominated by shallow Convolutional Neural Networks (CNNs). Although Transformer architectures have shown promise in other domains, they have yet to demonstrate stable performance improvements over CNNs in EEG tasks, primarily due to data scarcity. Furthermore, developing a foundation model for brain decoding capable of resolving these challenges remains far from practical applications.

The emergence of Vision-Language Models (VLMs) offers a promising alternative. Recall that clinical EEG analysis has often relied on visual inspection by experts for diagnostics, such as identifying epileptic seizures. Taking that inspiration, in this work, we propose a novel paradigm for EEG decoding by converting time-series signals into visual waveform plots and analyzing them using VLMs, whereas neuroscience domain expertise is integrated into textual prompts. To further condition the autoregressive outputs of VLMs, we propose a Retrieval-Augmented In-Context Learning (RAICL) framework. This approach dynamically selects the most representative and relevant EEG waveform plots as few-shot examples, serving as visual embeddings that contrast with or align with the test trial embeddings' patterns to condition the VLM's outputs, thereby decoding enhancing performance.

\begin{figure*}[htbp] \centering
\includegraphics[width=.9\linewidth,clip]{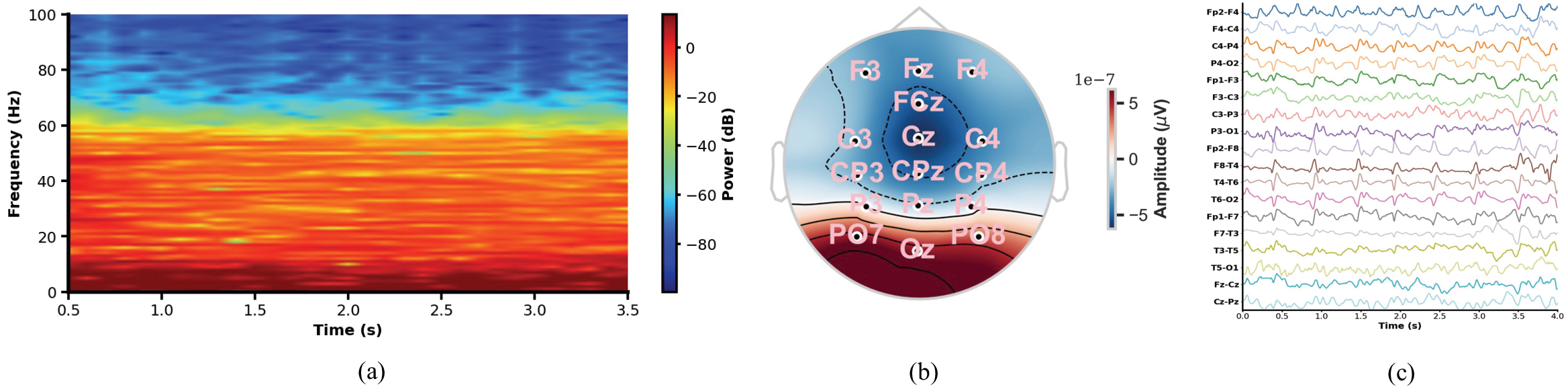}
\caption{Three types of EEG-to-image conversion methods. (a) time-frequency spectrogram, using a seizure EEG trial); (b) topographical map, using a P300 EEG trial; and (c) waveform plot, using a seizure EEG trial.} \label{fig:eegplots}
\end{figure*}

To our knowledge, this is the first work to explore EEG decoding using black-box VLMs, specifically targeting real-world consumer and clinical utility. Our main contributions are summarized as follows:
\begin{enumerate}
\item Demonstration of the efficacy of decoding EEG signals via waveform plots using VLMs, establishing a new paradigm for physiological signal analysis that bypasses the constraints of traditional signal-based learning.
\item Proposal of RAICL, a framework that retrieves semantically representative and relevant examples of EEG waveform plot to guide VLM generation, significantly improving decoding accuracy.
\item Comprehensive empirical analysis using seizure detection datasets, verifying that our approach, applicable to both API-based and open-source VLMs, achieves EEG decoding performance competitive with, or even superior to, state-of-the-art signal-based approaches.
\end{enumerate}

The remainder of this paper is organized as follows: Section~\ref{sect:relatedwork} reviews related work in EEG decoding and VLM applications. Section~\ref{sect:vlm4eegdecoding} details the methodology for EEG waveform plot decoding using VLMs under the RAICL framework. Section~\ref{sect:experiments} presents experimental results. Finally, Section~\ref{sect:conclusions} draws conclusions and outlines future research directions.

\section{Related Works}\label{sect:relatedwork}
This section reviews EEG decoding methodologies, ranging from classic signal processing to contemporary deep learning and image-based approaches.

\subsection{Signal-Based EEG Decoding}
Traditional EEG decoding pipelines typically involve signal filtering or heuristic feature extraction designed for specific experimental paradigms or tasks, followed by standard machine learning classifier or regressor. Examples include common spatial patterns for motor imagery sensorimotor function decoding~\cite{Blankertz2008}, xDAWN for event-related potential visual decoding~\cite{Rivet2009}, differential entropy for emotion recognition~\cite{Zheng2015}, and various statistical features for seizure detection~\cite{Wang2023TASA}.

Over the past decade, the decoding paradigm has been shifted to end-to-end deep learning, where raw signals are the inputs rather than hand-crafted features. CNNs architectures for EEG demonstrated good performance on small datasets, such as EEGNet~\cite{Lawhern2018EEGNet}, ShallowCNN, and DeepCNN~\cite{Schirrmeister2017}, facilitates unified spatio-temporal-spectral EEG decoding.

Transformer architectures such as Conformer~\cite{Song2023Conformer}, Deformer~\cite{Ding2024}, DBConformer~\cite{Wang2025a}, and ADFCNN~\cite{Tao2024} combine CNNs with attention mechanisms, resulting in deeper architectures with increased parameter density. These hybrid CNN-Transformer models have been shown to outperform pure CNNs when trained on sufficiently large datasets.

\subsection{Image-Based EEG Analysis}
Common techniques for converting EEG signals into visual formats include the following, as illustrated in Fig.~\ref{fig:eegplots}:
\begin{enumerate}
\item Time-Frequency Spectrogram: A 2D heatmap representing spectral energy distribution over time and frequency~\cite{Tabar2016, Mandhouj2021}. This is a temporal-spectral plot.
\item Topographical Map: A spatial projection of voltage or power onto a 2D circular plane within certain duration~\cite{Bashivan2016}. It can be viewed as a spatial plot.
\item Waveform Plot: Provides a continuous temporal fluctuation of electrical potentials at the scalp surface with time on the horizontal axis and voltage amplitude on the vertical axis~\cite{Emami2019}. It can be viewed as a temporal plot. When stacking signals of multiple channels onto one figure, it can be viewed as a spatial-temporal plot.
\end{enumerate}
Alternative conversion methods that do not strictly adhere to these patterns have also been proposed~\cite{Sun2025}.

\subsection{Vision-Language Models}
Large Language Models (LLMs)~\cite{Zhao2023a}, as autoregressive textual models, lack native perception of sensory modalities. Early attempts to encode time-series data directly as text tokens for LLMs~\cite{Hu2024, Zhao2025} suffer from an inherent modality gap and low information density per token.

VLMs~\cite{Zhang2024} align a visual encoder with an LLM's embedding space, facilitating the multimodal integration of visual and textual tokens for autoregressive generation. These models demonstrate a promising capability to transform EEG data into a modality compatible with Transformer-based architectures. By encoding EEG signals as images for VLMs, rather than text for LLMs, this approach not only enhances information density per token but also enables semantically rich and contextually relevant interaction via textual prompts.

Contemporary VLMs include both proprietary and open-source models. Proprietary systems, such as GPT-5 and Gemini-3~\cite{Team2025}, offer advanced reasoning capabilities via API but remain opaque regarding their internal architectures. Open-source alternatives like Qwen3-VL~\cite{Yang2025, Bai2025} and InternVL3~\cite{Zhu2025} have also achieved decent performance with parameters ranging from sub-billion to hundreds of billions.

\subsection{In-Context Learning}
In-Context Learning (ICL) enables models to adapt to new tasks by embedding input-output pairs directly into the inference prompt~\cite{Dong2024}. This technique conditions the model on the target label space and formatting requirements without parameter updates~\cite{Min2022}. Furthermore, strategies like Chain-of-Thought (CoT) enhance reasoning by incorporating intermediate logical steps into these exemplars~\cite{Wei2022}.

However, ICL performance is sensitive to example selection, necessitating the development of retrieval-augmented strategies to mitigate instability. In the textual domain, Khandelwal \emph{et al.}~\cite{Khandelwal2020} proposed retrieving training examples semantically closest to the test query in the feature space. Building on this, Ram \emph{et al.}~\cite{Ram2023} showed that prepending retrieved documents significantly reduces perplexity in language modeling. Additionally, Zhang \emph{et al.}~\cite{Zhang2022} utilized reinforcement learning to learn generalizable policies for optimal example selection.

For VLMs, selecting appropriate visual demonstrations is equally critical. Zhang \emph{et al.}~\cite{Zhang2023a} demonstrated that retrieving examples with similar semantic and spatial content yields superior conditioning for visual tasks. Zhou \emph{et al.}~\cite{Zhou2022} introduced context optimization, which optimizes learnable continuous vectors to maximize the alignment between visual and textual representations.

To our knowledge, the work most similar to ours is Qiu~\emph{et al.}~\cite{Qiu2025}, which predicts sleep stages using EEG waveform images using VLMs, but requires training a specialized vision encoder. In contrast, our RAICL approach works exclusively on the query, enabling zero-training deployment compatible with high-performance, black-box, proprietary VLMs.

\section{Method}\label{sect:vlm4eegdecoding}
\begin{figure*}[htpb] \centering
\includegraphics[width=1.02\linewidth,clip]{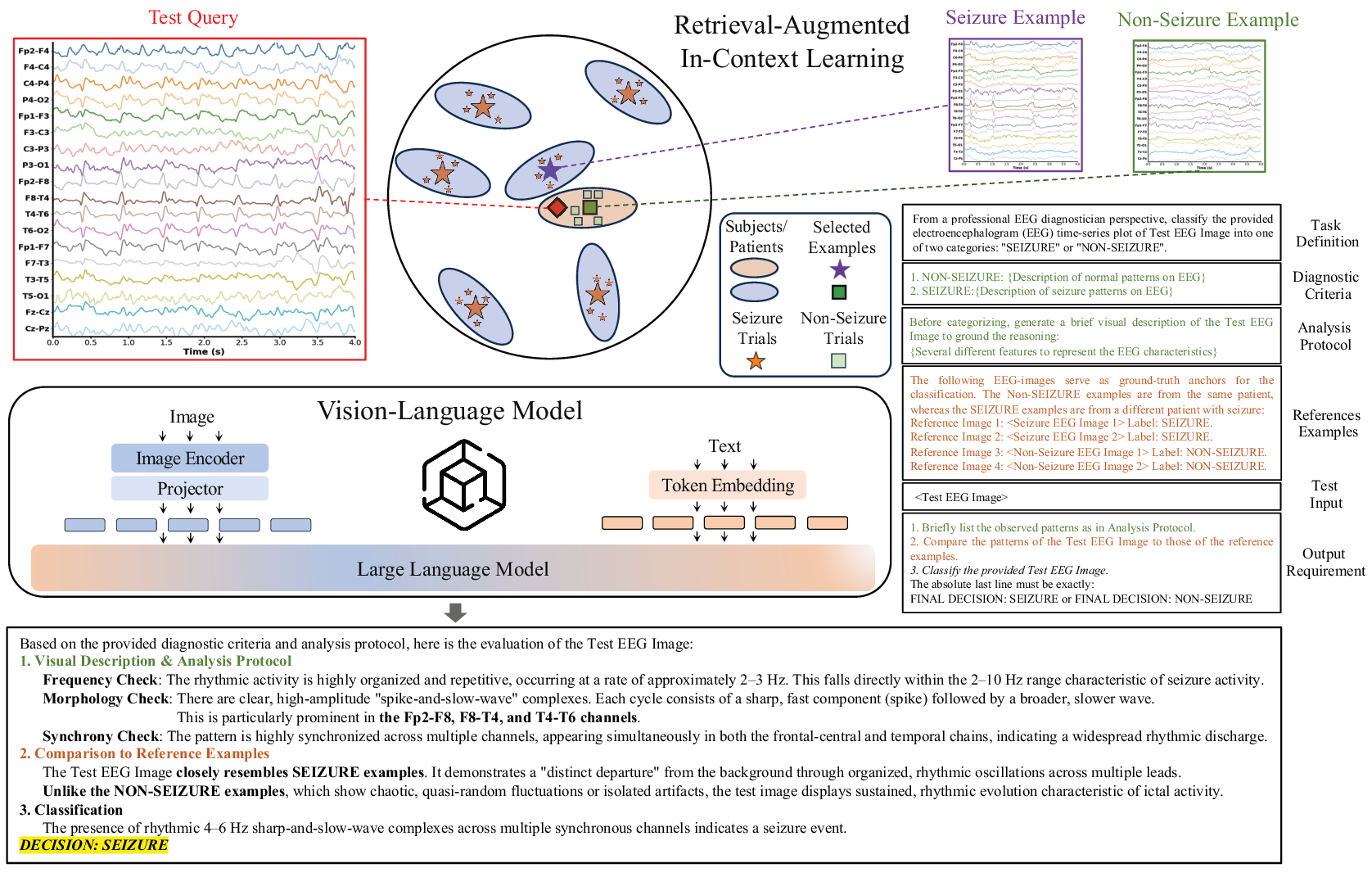}
\caption{Overview of the VLM for EEG waveform decoding with RAICL pipeline. The multi-variate EEG time-series is first rendered into a stacked waveform image, and then encoded into high-dimensional embeddings via an image encoder. Certain VLM architectures utilize a projector to map these embeddings into the language model's token space. Native multimodal models treat visual and textual tokens identically within a unified embedding space. For the textual prompt, domain expertise is injected as textual tokens via a CoT framework, as indicated in green. Few-shot examples facilitate ICL in the form of visual tokens, as indicated in red. RAICL strategies introduce metrics of representativeness and similarity in example selection. Finally, the visual and textual token embeddings are concatenated to form a multimodal prompt, querying the language model for autoregressive reasoning and inference.}
\label{fig:vlm4eeg}
\end{figure*}

We propose a decoding framework that reformulates multi-channel EEG analysis as a multimodal query to an off-the-shelf VLM. This approach eliminates the need for parameter updates, thereby facilitating rapid deployment in clinical and consumer scenarios. Illustrated in Fig.~\ref{fig:vlm4eeg}, the framework addresses three primary challenges:
\begin{enumerate}
\item Visual Encoding: The conversion of EEG trials into waveform images optimized for precise embedding by a visual encoder.
\item Prompt Engineering: The construction of textual prompts that elicit effective reasoning.
\item Example Retrieval: The augmentation of multimodal prompts with selected support set examples.
\end{enumerate}

\subsection{EEG Waveform Plot as Visual Tokens}\label{sect:eegwaveformplot}
The efficacy of a VLM in physiological signal decoding fundamentally relies on the fidelity of the visual embedding. Unlike natural images, EEG's multivariate spatio-temporal-spectral dynamics that must be encoded in an appropriate format to yield representations compatible with computer vision encoders.

The preprocessed EEG time-series $X \in \mathbb{R}^{C \times T}$ (where $C$ denotes the number of channels and $T$ denotes time samples) is transformed into a multi-channel stacked waveform plot $I_{\text{test}}$, which is subsequently forwarded to the VLM's visual encoder. This rendering pipeline incorporates three critical design choices, with details followed.

\textbf{Vertical Waveform Stacking} To visualize multivariate EEG signals while prioritizing information density, the amplitude $x_{c,t}$ of channel $c$ at time $t$ is mapped to a vertical image coordinate $y_{c,t}$ via a linear transformation:
\begin{align}
y_{c,t} = \alpha \cdot \text{Norm}(x_{c,t}) + \delta \cdot c, \label{eq:waveform}
\end{align}
where $\text{Norm}(\cdot)$ is a robust scaling function (e.g., median absolute deviation) used to standardize amplitudes, $\alpha$ is a scale hyperparameter, and $\delta$ is a vertical offset hyperparameter. This layout preserves temporal synchrony, allowing the VLM to scan vertical slices of the image to identify global or hemispheric patterns.

\textbf{Chromatic Encoding.} Signal amplitudes often exhibit significant variance, leading to potential overlap between adjacent channels. To mitigate ambiguity, we assign distinct colors to each channel using a high-contrast qualitative colormap. This chromatic encoding serves to visually disentangle the signals. In standard grayscale plots, spatially adjacent waveforms may become indistinguishable during high-amplitude events (e.g., F7-T3 and T3-T5 in Fig.~\ref{fig:vlm4eeg}). By leveraging the VLM's sensitivity to chromatic variation in RGB space, crossing trajectories can be effectively segregated. To maximize separability, adjacent channels can be assigned colors that maximize discriminability in the RGB embedding space (e.g., utilizing orthogonal vectors such as (0, 255, 255), (255, 0, 0), and (122, 122, 122)), rather than relying solely on perceptual distinctness for human observers. Alternatively, amplitude clipping could also handle overlap, but is suboptimal as it compromises signal integrity.

\textbf{High-Fidelity Rendering} Fine-grained morphological features, such as transient high-frequency oscillations or sharp waveform inflections, are critical diagnostic events. Standard plotting libraries often produce low-resolution bitmaps that alias these high-frequency components into artifacts, particularly when resized to the computer vision model's native input resolution (e.g., $224 \times 224$ pixels). We mitigate this by employing thickened line strokes and removing non-informative visual elements (e.g., grid lines and margins). This ensures that the tokenization process preserves essential geometric properties, facilitating both explicit optimal character recognition and implicit visual embedding comprehension.

\subsection{Chain-of-Thought Prior Expertise with Textual Tokens}
Although general-purpose foundation models possess broad knowledge, they often lack the specific expertise required to interpret complex physiological signals and its related task patterns. To bridge this semantic gap, domain-specific expertise are injected directly into the linguistic prompt, establishing a logical framework for the query.

The Chain-of-Thought (CoT) prompt design principle~\cite{Wei2022} is adopted, which shifts the task from straightforward question answering to structured reasoning. The prompt consists of the following specific components for EEG decoding:
\begin{enumerate}
\item Diagnostic Criteria: Explicit definitions of characteristics distinguishing patterns of different classes. This anchors the model's attention to relevant features (e.g., frequency bands, synchronization) rather than visual artifacts.
\item Analysis Protocol: A step-by-step reasoning guide compelling the model to generate intermediate descriptors before concluding. This enforces a clear separation between perception and decision-making.
\item Output Constraints: Strict formatting requirements for the response. This ensures the output is parsable for decision extraction and transparent in its rationale, thereby enhancing explainability.
\end{enumerate}

\subsection{Retrieval-Augmented In-Context Learning} \label{sect:raicl}
To further enhance decoding performance, we adopt the ICL design~\cite{Dong2024}, which conditions the model via labeled few-shot examples. Specifically, we prepend a support set of labeled EEG waveform images to the test query.

Let $\mathcal{S}$ denote a support set constructed for $M$-shot ICL, containing $M$ example pairs for each of the $K$ classes $\mathcal{S} = \{ (I_{\text{sup}}^{1}, y^{1}), \dots, (I_{\text{sup}}^{M \times K}, y^{M \times K}) \}$, where $y$ represents the class label as a textual token. Given a test EEG image $I_{\text{test}}$, the VLM approximates the conditional distribution $P(y \mid I_{\text{test}}, \mathcal{S})$ via autoregressive generation based on the concatenated multimodal context.

Since ICL performance is highly sensitive to $\mathcal{S}$, it necessitates curating the example selection process instead of randomly picking samples.

\subsubsection{Problem Formulation of RAICL}
A cross-subject, task-zero-shot setting is considered,  which reflects realistic clinical and consumer EEG application deployment scenarios. The test subject is assumed to possess a pool of historical non-task recordings (e.g., resting-state trials, which can be easily acquired during pre-use calibration) but no labeled positive (task) trials. However, labeled samples for both task and non-task states are available from a set of auxiliary subjects.

Formally, let $l=\{1,\dots,L\}$ index $L$ auxiliary source subjects. The $l$-th source subject possesses $n_{s,l}$ labeled EEG trials $\mathcal{D}_{s,l} = \{(I_{s,l}^i, y_{s,l}^i)\}_{i=1}^{n_{s,l}}$, where $y \in \{0, ..., K-1\}$. The test subject possesses a historical pool of non-task recordings $\mathcal{D}_{t} = \{(I_{t}^j, y_{t}^j)\}_{j=1}^{n_{t}}$, where $y_{t}^j=0$ for all $j$.

Let $f(\cdot)$ denote the fixed visual embedding function of the VLM's encoder. In a black-box API context, this corresponds to any off-the-shelf encoders of VLMs such as CLIP~\cite{Radford2021} or Qwen-VL~\cite{Bai2025}. For any test query $I_{\text{test}}$, our objective is to construct an optimal support set $\mathcal{S}$ containing $M$ demonstrations per class.

\subsubsection{Representativeness}
EEG signals are characterized by significant non-stationarity, where background brain activity varies substantially due to factors such as mental state, artifacts, movements, etc. Consequently, accurate decoding requires differentiating task-induced patterns from a subject-specific resting-state baseline. To minimize such distributional shift, non-task anchors are strictly selected from the test subject's own historical pool, $\mathcal{D}_{t}$, to establish anchors of non-task baselines.

To ensure these anchors are prototypically representative and robust to outliers, we employ a centroid-based selection strategy. First, we compute the test subject's non-task centroid in the embedding space:
\begin{align}
\mathbf{c}_{t}^{0} = \frac{1}{n_t} \sum_{j=1}^{n_t} f(I_{t}^j).\label{eq:centroid_ns}
\end{align}

We define the distance metric $d(\cdot, \cdot)$ as the cosine distance:
\begin{align}
d(\mathbf{u}, \mathbf{v}) = 1 - \frac{\mathbf{u}^{\top}\mathbf{v}}{\|\mathbf{u}\|_2 \|\mathbf{v}\|_2}.\label{eq:cosdist}
\end{align}

We then define the representativeness metric, $m_{\text{rep}}$, as the distance to its class centroid:
\begin{align}
m_{\text{rep}}(I_t) = d\big(f(I_t), \mathbf{c}_{t}^{0}\big). \label{eq:repmetric}
\end{align}

The non-task support set $\mathcal{S}_{0}$ is constructed by selecting the $M$ trials from $\mathcal{D}_{t}$ that minimize $m_{\text{rep}}$ (i.e., the medoids closest to the mean representation of the subject's resting-state trials).

\subsubsection{Similarity}
Demonstrations for task classes $k \in \{1, \dots, K-1\}$ must be retrieved from auxiliary source subjects, as the test subject is assumed to be zero-shot for these states. To mitigate outlier selection and reduce computational overhead, we employ a nearest medoid strategy. This approach restricts the search space to high-quality subject-specific prototypes rather than the entire auxiliary dataset, and requires the example to be the most similar to the test query.

For each auxiliary subject $l$ and task class $k$, we identify a single representative trial, the medoid $I_{s,l}^{\text{medoid}}$. This is defined as the instance within the auxiliary subject's trial set $S_{k}^{(s,l)}$ that minimizes the cosine distance to the set's arithmetic mean.

We then construct the task support set by comparing the test query $I_{\text{test}}$ against this curated pool of medoids, $\mathcal{D}_{s,l}^{\text{medoids}}=\{ I_{s,1}^{\text{medoid}}, \dots, I_{s,L}^{\text{medoid}} \}$. The relevance of each prototype is scored based on its similarity to the query of representation:
\begin{align}
m_{\text{sim}}(I_{s,l}^{\text{medoid}}) = d\big(f(I_{s,l}^{\text{medoid}}), f(I_{\text{test}})\big).\label{eq:repsim}
\end{align}
The $M$ medoids with the highest similarity (lowest distance) are selected for the support set. This strategy ensures that retrieved examples are both valid class representatives and morphologically similar to the specific test trial.

The complete procedure is summarized in Algorithm~\ref{alg:raicl}. The complete algorithm establishes a simple, straightforward, yet robust pipeline for reliable EEG decoding via VLMs.

\begin{algorithm}[htpb]
\caption{Vision-Language Model for EEG Decoding via Retrieval-Augmented In-Context Learning (RAICL).}  \label{alg:raicl}
\begin{algorithmic}
\REQUIRE $X \in \mathbb{R}^{C \times T}$, raw EEG test trial;\\
$\mathcal{D}_t = \{(I_{t}^j, 0)\}_{j=1}^{n_t}$, non-task resting-state samples from test subject;\\
$\mathcal{D}_{s,l} = \{(I_{s,l}^i, y_{s,l}^i)\}_{i=1}^{n_{s,l}}$, auxiliary datasets from $L$ source subjects;\\
$f(\cdot)$, a visual encoder;\\
$M$, number of examples per class;\\
$K$, total number of classes;\\
\ENSURE Multimodal prompt $\mathcal{P}$.
\STATE Convert raw signal $X$ to image $I_{\text{test}}$ by (\ref{eq:waveform});
\STATE Encode test EEG embedding $f(I_{\text{test}})$;
\STATE Initialize support set $\mathcal{S} = \emptyset$;
\STATE Calculate history centroid $\mathbf{c}_t^0$ using $\mathcal{D}_t$ by (\ref{eq:centroid_ns});
\STATE Calculate representativeness $m_{\text{rep}}(I_t)$ for all $I_t \in \mathcal{D}_t$ by (\ref{eq:cosdist}) and (\ref{eq:repmetric});
\STATE Construct non-task support set $\mathcal{S}_0$ by selecting $M$ trials with lowest $m_{\text{rep}}$;
\STATE Update $\mathcal{S} \leftarrow \mathcal{S} \cup \mathcal{S}_0$;
\FOR{$k=1$ to $K-1$}
	\STATE Initialize class medoid pool $\mathcal{M}^{k} \leftarrow \emptyset$;
    \FOR{$l=1$ to $L$}
        \STATE Compute centroid $\mathbf{c}_{s,l}^k$ for source subject $l$, class $k$;
        \STATE Identify medoid $I_{s,l}^{\text{medoid}}$ by minimizing distance to $\mathbf{c}_{s,l}^k$, calculated similarly by (\ref{eq:centroid_ns});
        \STATE Update $\mathcal{M}^{k} \leftarrow \mathcal{M}^{k} \cup \{ I_{s,l}^{\text{medoid}} \}$;
    \ENDFOR
    \STATE Calculate query distance $m_{\text{dist}}(I)$ for all $I \in \mathcal{M}^{k}$ relative to $I_{\text{test}}$ by (\ref{eq:repsim});
    \STATE Construct task support set $\mathcal{S}_k$ by selecting $M$ medoids with lowest $m_{\text{dist}}$;
    \STATE Update $\mathcal{S} \leftarrow \mathcal{S} \cup \mathcal{S}_k$;
\ENDFOR
\STATE Construct multimodal prompt $\mathcal{P}$ by concatenating $\mathcal{S}$, $I_{\text{test}}$, and textual instructions;
\end{algorithmic}
\end{algorithm}

\section{Experiments}\label{sect:experiments}
This section presents the empirical validation on seizure detection tasks. Code used, query prompts, and model responses are publicly available\footnote{\url{https://github.com/sylyoung/VLM4EEG}}.

\subsection{Datasets and Settings}
Two public seizure EEG datasets were used:
\begin{enumerate}
\item Dataset collected at Wuhan Children’s Hospital affiliated to Tongji Medical College of the Huazhong University of Science and Technology, referred to as CHSZ~\cite{Peng2022}. EEG data were recorded from infant and children patients. Each patient had one to six seizure events. Experts annotated the beginning and end of each seizure for each child.
\item Dataset collected at the Helsinki University Hospital neonatal intensive care unit from neonatal, referred to as NICU~\cite{Stevenson2019}. EEG recorded neonatal seizures from 79 full-term neonates, with median recording duration being 74 minutes. Three experts annotated every second of EEG data independently. According to the consensus of annotations, 39 neonates had seizures and were retained.
\end{enumerate}
Table~\ref{tab:seizuredatasets} summarizes their main characteristics. Proprocessing steps follow our previous publication to split into 4-second non-overlapping trials with bipolar channels~\cite{Wang2023TASA}.

\begin{table}[htpb] \centering \setlength{\tabcolsep}{1mm}
\caption{Summary of the statistics of the two seizure EEG datasets.}
\begin{tabular}{c|c|c|c|c|c|c}
\toprule
\multirow{2}{*}{Dataset} & \multirow{2}{*}{\# Patients} & \# Channels & \# Total & \# Seizure & Trial & Sampling \\
& & (bipolar) & Trials & Trials & Duration (s) & Rate (Hz) \\ \midrule
CHSZ & 27 & 18 & 21,237 & 716 & 4 & 250 \\
NICU & 39 & 18 & 52,534 & 11912 & 4 & 256 \\
\bottomrule
\end{tabular}\label{tab:seizuredatasets}
\end{table}

For baseline signal-based approaches, we used a standard leave-one-subject-out cross-validation. In this scheme, one subject is held out for testing, whereas the remaining subjects constitute the training set. This strict separation prevents the temporal information leakage often observed in studies reporting inflated performance metrics~\cite{Kapoor2023}, ensuring a rigorous assessment of generalization performance of decoding models.

For the VLM framework, we simulated a realistic zero-calibration scenario. To strictly prevent temporal leakage, the non-task (resting-state) anchors used for the few-shot support set were selected exclusively from the test subject's historical pool (i.e., trials occurring prior to the test query). Positive task trials (seizures) from the test subject were excluded from the support set, as described in Section~\ref{sect:raicl}.

To manage computational overhead and query volume, datasets were downsampled prior to VLM inference. For CHSZ, we retained every tenth non-seizure trial. For NICU, we retained every tenth trial for all classes.

\subsection{Decoding Models}

For signal-based approaches, the raw EEG signal are the direct input. Models were trained from scratch following a standard cross-subject decoding setting, using EEG from auxiliary subjects as training data and holding out the test subject. We compared the following architectures:
\begin{enumerate}
\item ShallowCNN~\cite{Schirrmeister2017}: A two-layer CNN optimized for temporal and spatial filtering.
\item DeepCNN~\cite{Schirrmeister2017}: A five-layer CNN architecture following a design similar to standard computer vision models but adapted for EEG.
\item EEGNet~\cite{Lawhern2018EEGNet}: A compact CNN using depthwise and separable convolutions to minimize trainable parameters.
\item Conformer~\cite{Song2023Conformer}: A hybrid CNN-Transformer architecture where the CNN part mimics ShallowCNN and Transformer blocks captures long-range temporal dependencies.
\item Deformer~\cite{Ding2024}: A CNN-Tranformer architecture that further enhances temporal dynamics extraction.
\item ADFCNN~\cite{Tao2024}: A dual-branch CNN model that fuses features from independent temporal and spatial branches via an attention mechanism.
\item DBConformer~\cite{Wang2025a}: A parallel dual-branch architecture that explicitly decouples and combines temporal and spatial representation learning.
\end{enumerate}

For image-based approaches, the stacked waveform image is the input. We initialized these models with pre-trained weights from ImageNet-1K and fine-tuned them on auxiliary subjects' data.
\begin{enumerate}
\item GoogLeNet~\cite{Szegedy2015}: Introduces parallel convolutions to capture multi-scale features while remaining computationally efficient.
\item ResNet~\cite{He2016}: A classic architecture that employs residual skip connections to mitigating the vanishing gradient problem.
\item DenseNet~\cite{Huang2017}: Features a dense connectivity pattern where each layer receives the feature maps of all preceding layers as input, maximizing feature reuse.
\item EfficientNet~\cite{Tan2019}: Employs a compound scaling method that uniformly scales depth, width, and resolution using a principled coefficient.
\item SwinTransformer~\cite{Liu2021}: A hierarchical Vision Transformer that computes self-attention within local, non-overlapping windows and uses a shifted windowing scheme to enable cross-window communication.
\item ViT~\cite{Dosovitskiy2021}: Abandons convolutions entirely, applying a pure self-attention mechanism to a sequence of flattened image patches.
\end{enumerate}

For VLMs, we used Gemini-3-Flash\footnote{\url{https://deepmind.google/models/gemini/flash/}.}~\cite{Team2025}, which is a closed-source state-of-the-art VLM that is available via API. We also compared to open-source VLMs of Qwen3-VL-32B-Instruct\footnote{\url{https://huggingface.co/Qwen/Qwen3-VL-32B-Instruct}.}~\cite{Yang2025, Bai2025} and InternVL3-38B\footnote{\url{https://huggingface.co/OpenGVLab/InternVL3-38B}.}~\cite{Zhu2025} that are locally deployed with weights from Huggingface~\cite{Wolf2019}, which represent the open-source state-of-the-art VLMs. To eliminate stochasticity in the model's predictive output, the sampling temperature was set to $0$.

The $f(\cdot)$ of RAICL used the pre-trained CLIP~\cite{Radford2021} visual encoder, ensuring fairness comparison of the effect of RAICL for all VLMs tested. We used $M=2$ examples per class in all experiments. Balanced classification accuracy (BCA) was used as the metric for evaluation, considering the class-imbalance that existed in the datasets.

\subsection{Results on Seizure Detection} \label{sect:seizureresults}

The main seizure detection results are summarized in Table~\ref{tab:seizure}. Given the large number of patients in the experiments, Table~\ref{tab:seizure} reports performance averaged across all patients.

Observe that:
\begin{enumerate}
\item Task-zero-shot EEG-based seizure detection remains highly challenging. Without access to patient-specific seizure records, detection performance is far from perfect.
\item Among signal-based decoding models, CNN-based architectures consistently achieved reliable performance, whereas Transformer-based models did not demonstrate clear improvements over CNNs.
\item Notably, pure vision-based decoding using EEG waveform images achieved strong performance. In particular, the ResNet model operating on EEG waveform images outperformed all CNN and Transformer architectures specifically designed for EEG analysis. This result validates the effectiveness of the proposed EEG waveform image representation and suggests that, given sufficient fine-tuning data, EEG signals can be effectively decoded using pre-trained computer vision models.
\item In contrast, the ViT model exhibited inferior performance, indicating that their highly global and flexible attention mechanisms may be ill-suited for the proposde EEG stacked waveform image. CNNs capture short-term temporal dependencies, local cross-channel correlations more effectively, being more robust to minor temporal misalignments. SwinTransformer likely benefits from controlled locality through windowed and shifted self-attention mechanisms.
\item Leading open-source VLMs such as InternVL and QwenVL still could not reach performance as that of Gemini of proprietary ones. As Gemini is a native multimodal model, it is likely the vision encoder rather than the language model that matters more for this task.
\item Under the proposed RAICL framework, the Gemini-3-Flash VLM outperformed all existing signal-based and image-based approaches. By incorporating textual prior knowledge and informed example selection, the model overcomes the limited capacity of traditional signal-based models and leverages complementary expertise from the textual modality.
\end{enumerate}

\begin{table}[htpb] \centering \setlength{\tabcolsep}{1.1mm}
\caption{Average BCA (\%) in EEG-based seizure detection over all patients. No labeled task (i.e., seizure) class trials from the test patient are accessible by the decoding models.}
\label{tab:seizure}
\begin{tabular}{c|c|c|c}
\toprule
Input Type & Model & CHSZ & NICU \\
\midrule
\multirow{7}{*}{\shortstack[c]{Raw \\ EEG \\ Signal}} & ShallowCNN & 79.47$_{\pm0.70}$ & 61.68$_{\pm0.85}$ \\
& DeepCNN & 80.38$_{\pm1.36}$ & 57.44$_{\pm0.53}$ \\
& EEGNet & 80.65$_{\pm0.92}$ & 65.32$_{\pm0.60}$ \\
& Conformer & 75.62$_{\pm1.21}$ & 64.51$_{\pm0.49}$ \\
& Deformer & 75.26$_{\pm1.45}$ & 59.28$_{\pm0.57}$ \\
& ADFCNN & 81.03$_{\pm0.55}$ & 63.75$_{\pm0.63}$ \\
& DBConformer & 80.98$_{\pm0.44}$ & 65.70$_{\pm0.69}$ \\
\midrule
\multirow{6}{*}{\shortstack[c]{EEG \\ Waveform \\ Image}} & GoogleNet & 77.07$_{\pm0.43}$ & 66.15$_{\pm1.24}$ \\
& ResNet & 81.80$_{\pm0.17}$ & 69.98$_{\pm1.81}$ \\
& DenseNet & 81.81$_{\pm0.31}$ & 69.71$_{\pm2.13}$ \\
& EfficientNet & 81.72$_{\pm0.24}$ & 68.03$_{\pm1.61}$ \\
& SwinTransformer & 81.48$_{\pm0.14}$ & 68.95$_{\pm1.24}$ \\
& ViT & 75.89$_{\pm0.25}$ & 67.41$_{\pm1.48}$ \\
\midrule
\multirow{3}{*}{\shortstack[c]{EEG \\ Waveform \\ Image \& Prompt}} & InternVL3-38B & 74.91 & 60.90 \\
& Qwen3-VL-32B-Instruct & 75.84 & 66.01 \\
& Gemini-3-Flash & \textbf{82.10} & \textbf{70.08} \\
\bottomrule
\end{tabular}
\end{table}

\subsection{Effectiveness of CoT and ICL} \label{sect:textualablation}

\begin{figure}[htpb] \centering
\includegraphics[width=\linewidth,clip]{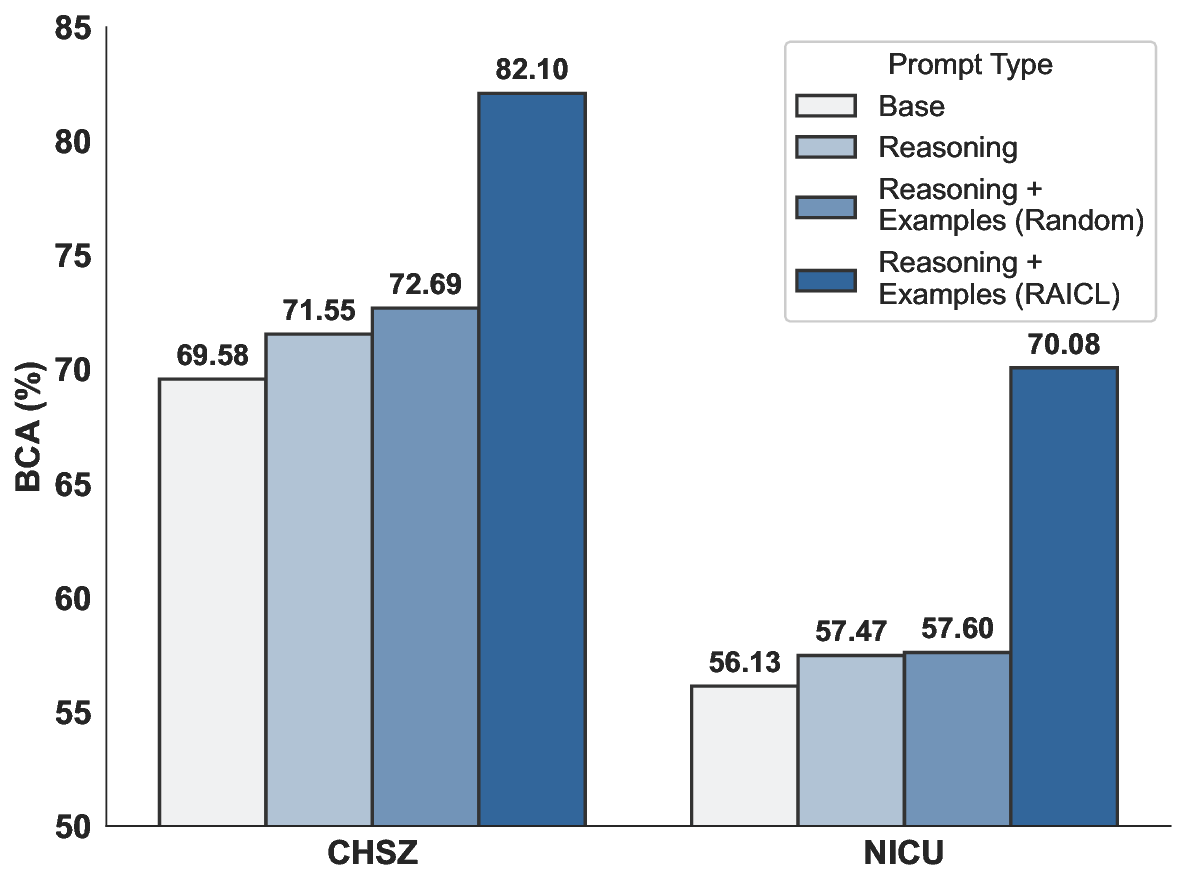}
\caption{Ablation studies of prompt design on two seizure EEG datasets.} \label{fig:bar}
\end{figure}

Fig.~\ref{fig:bar} presents ablation studies evaluating the impact of incorporating textual prior knowledge via CoT prompting and few-shot ICL. Specifically, three prompt configurations were considered:
\begin{enumerate}
\item Base: consists solely of the black text shown in Fig.~\ref{fig:vlm4eeg}, directly querying the model for a classification decision.
\item Reasoning: includes the black and green text in Fig.~\ref{fig:vlm4eeg}, augmenting the query with prior expert knowledge describing class-specific EEG patterns.
\item Reasoning + Examples (Random): includes the full prompt of black, green, and red text in Fig.~\ref{fig:vlm4eeg}, incorporating both prior expertise and randomly selected few-shot examples.
\end{enumerate}
Observe that CoT and ICL both offered performance improvement. However, the gains obtained from randomly selected few-shot examples are relatively modest, motivating the need for more principled example selection strategies, as addressed by the proposed RAICL framework. The RAICL selection strategies yielded significant performance improvements over random selection, with average BCA gains exceeding 10\% under more refined selection strategies.

\subsection{Effectiveness of Selection Strategies of RAICL} \label{sect:visualablation}

\begin{figure*}[htpb] \centering
\subfigure[]{\includegraphics[width=.49\linewidth,clip]{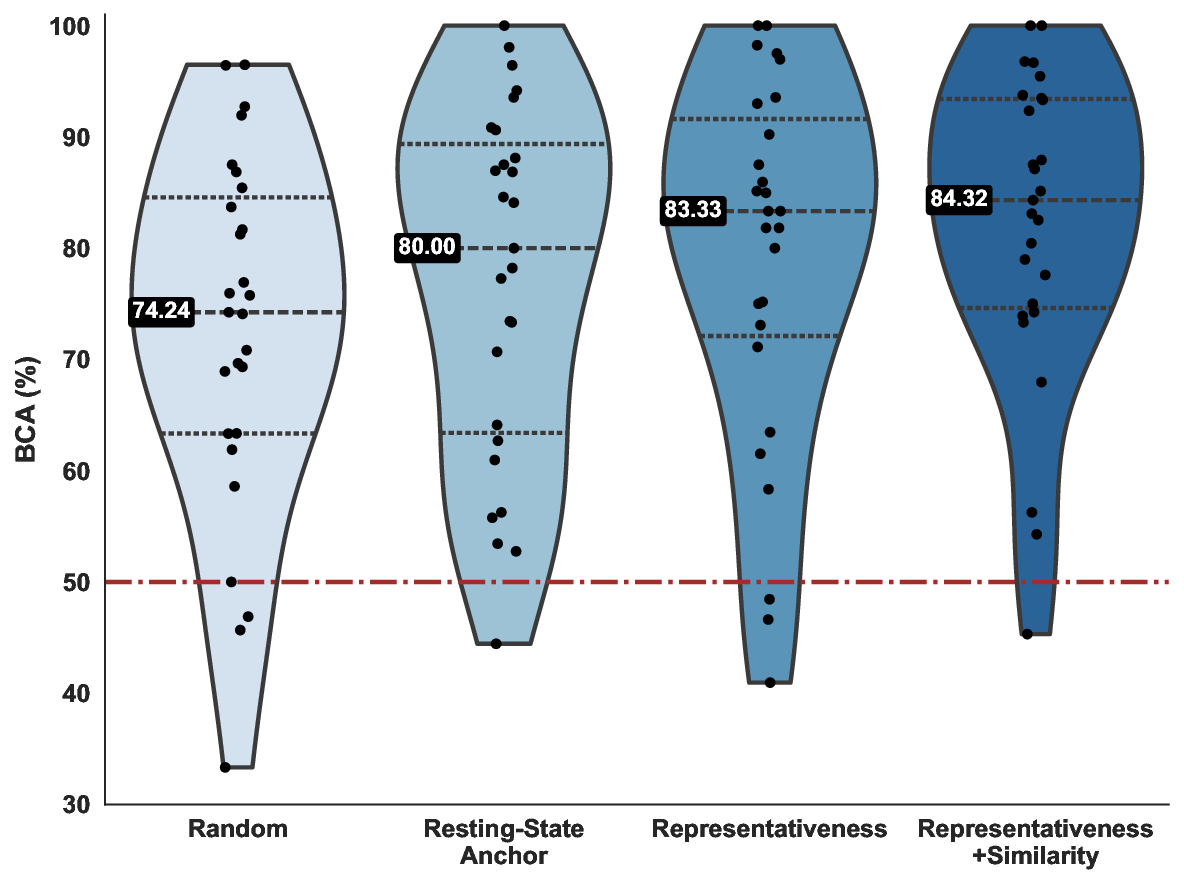}\label{fig:CHSZ-violin}}
\subfigure[]{\includegraphics[width=.49\linewidth,clip]{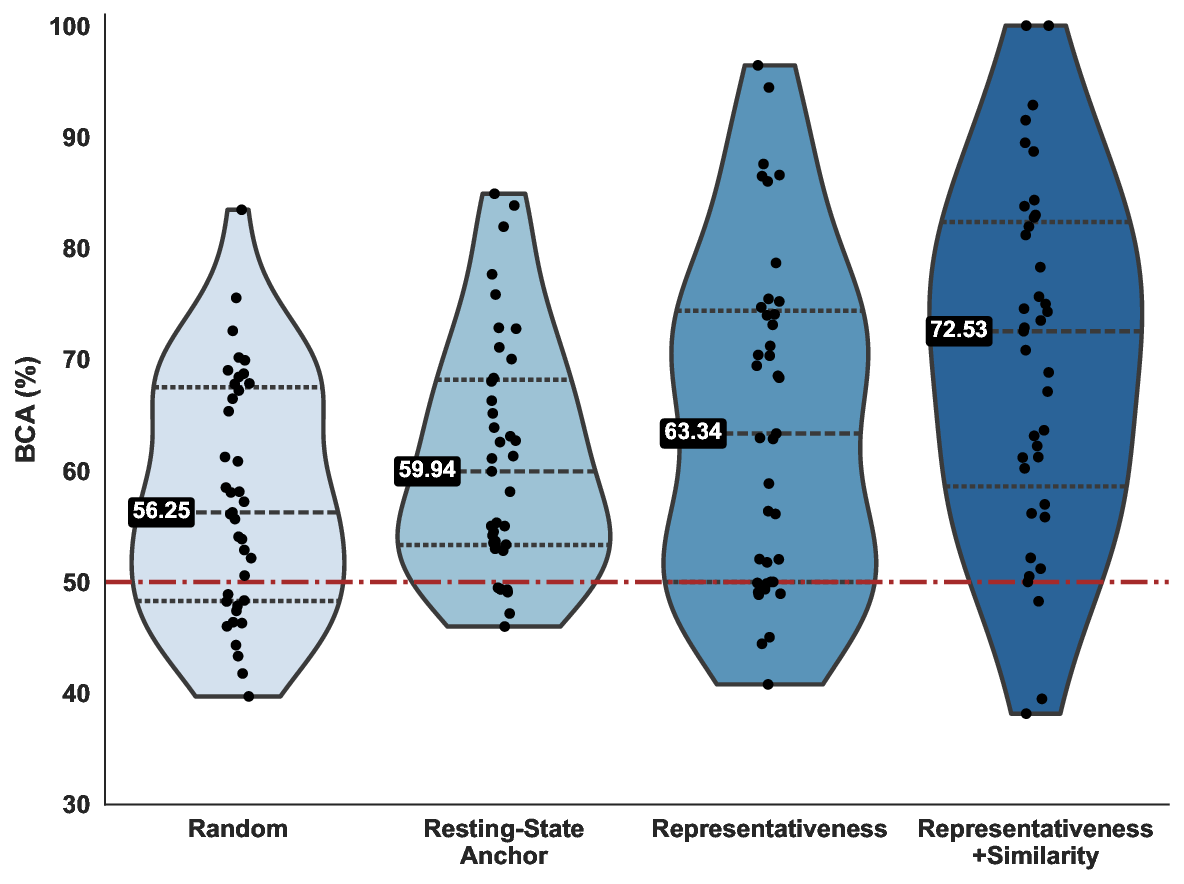}\label{fig:NICU-violin}}
\caption{Patient-wise performance in EEG-based seizure detection using Gemini-3-Flash. Ablation studies are conducted on selection strategies of RAICL for the two-shot examples. Each dot denotes the performance for a patient. The red dashed line indicates chance-level performance, and the gray dashed lines in the violin denote quartiles/median. (a) CHSZ dataset; and (b) NICU dataset.} \label{fig:violin}
\end{figure*}

To further assess the effectiveness of RAICL selection strategies, Fig.~\ref{fig:violin} reports patient-wise performance along with ablation results for different example selection mechanisms. The following strategies were evaluated:
\begin{enumerate}
\item Random: few-shot examples are randomly selected from auxiliary patients.
\item Resting-State Anchor: non-seizure examples are randomly sampled from the test patient’s historical recordings, whereas seizure examples are randomly selected from auxiliary patients.
\item Representativeness: non-seizure examples are selected as medoids from the test patient’s historical recordings, and seizure examples are selected from the medoids from auxiliary patients.
\item Representativeness + Similarity: non-seizure examples are selected as medoids from the test patient’s historical recordings, and seizure examples are chosen as the most similar seizure medoids from auxiliary patients with respect to the test trial.
\end{enumerate}
As shown in Fig.~\ref{fig:violin} of violin plot, each strategy of RAICL offered noticeable performance gains. These results highlight both the necessity of ICL and the effectiveness of the proposed RAICL selection mechanisms.

\begin{figure*}[htpb] \centering
\subfigure[]{\includegraphics[width=.49\linewidth,clip]{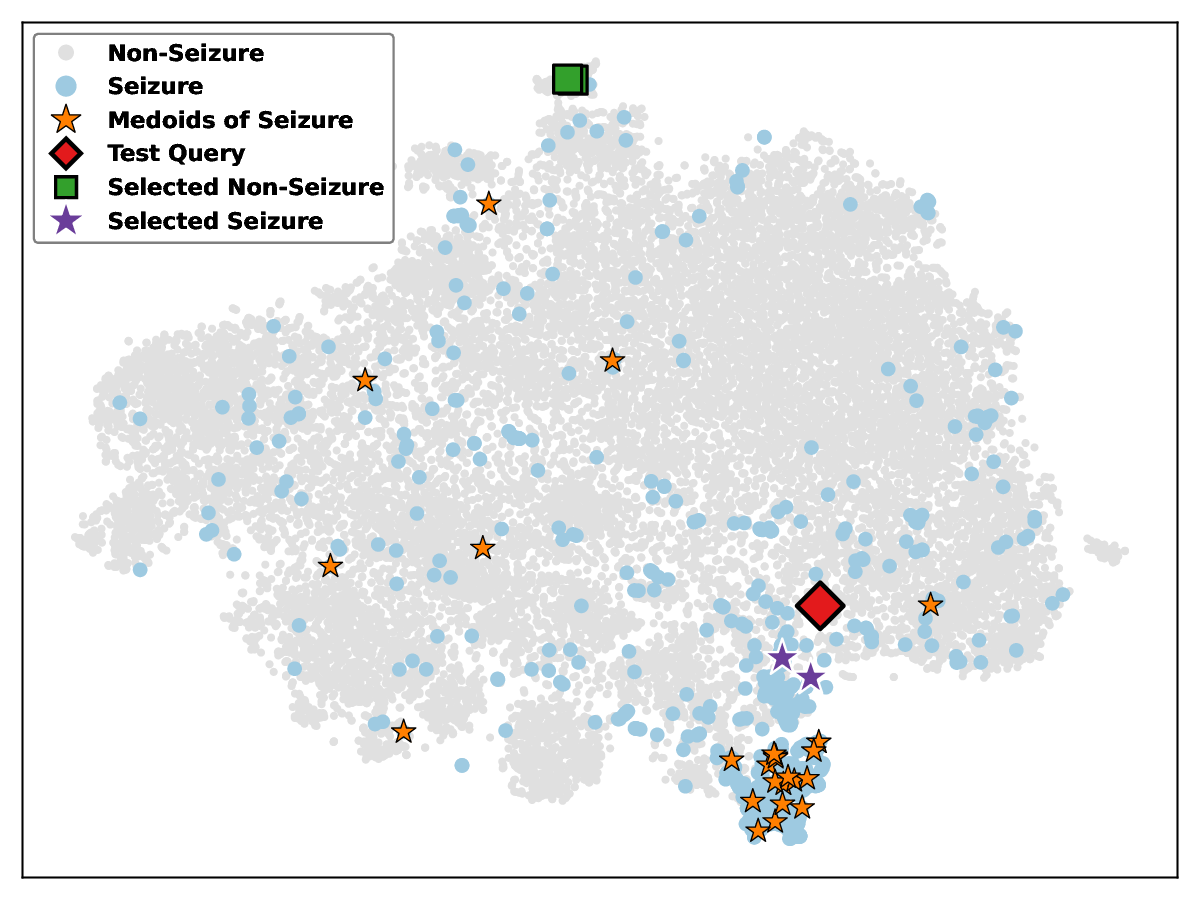}\label{fig:SEIZURE-REP+SIM}}
\subfigure[]{\includegraphics[width=.49\linewidth,clip]{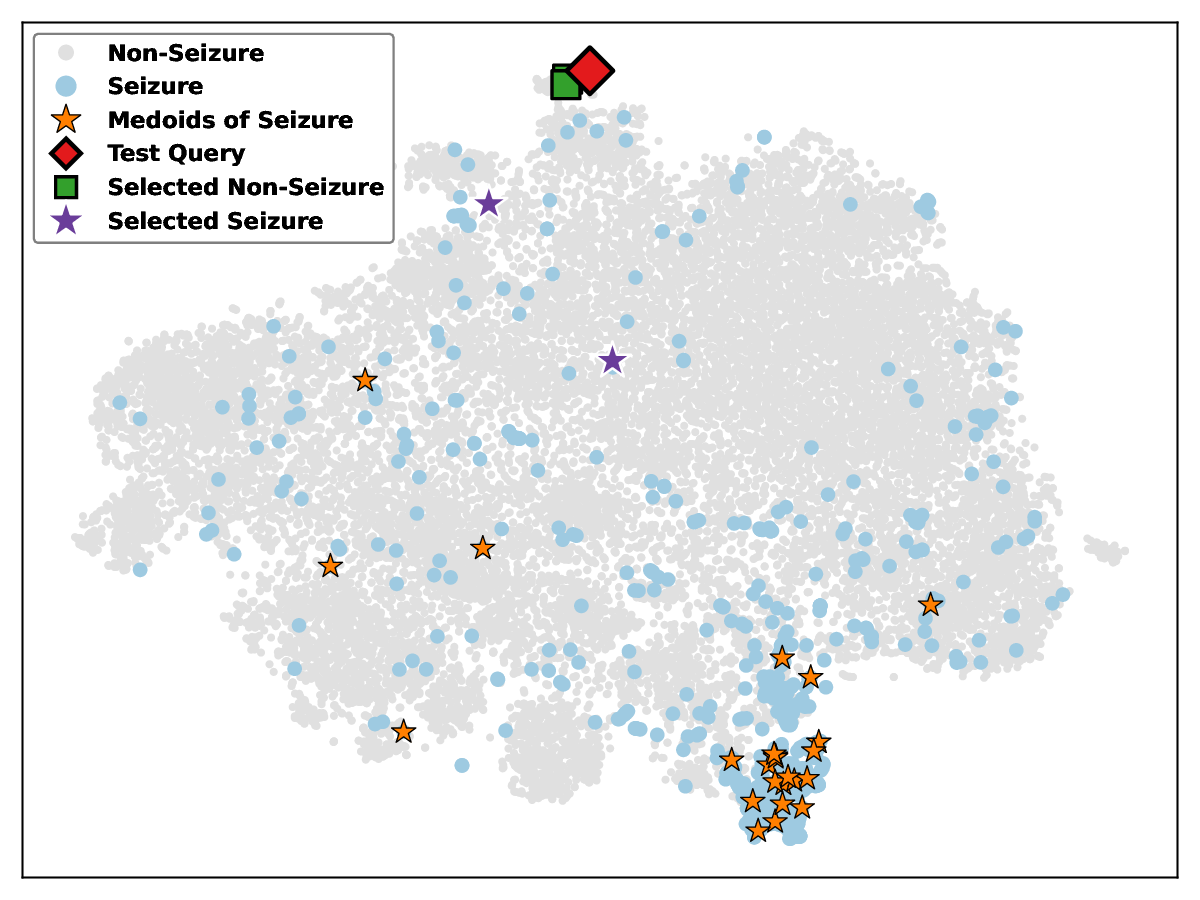}\label{fig:Non-SEIZURE-REP+SIM}}
\caption{$t$-SNE visualization of RAICL selection strategies, with both representativeness + similarity, given $M=2$ (two-shot) examples. Embeddings of EEG waveform images are extracted using the visual encoder on the CHSZ dataset. (a) Seizure test query; (b) Non-seizure test query.} \label{fig:tsne}
\end{figure*}

Fig.~\ref{fig:tsne} visualizes the behavior of RAICL selection strategies. When the test query is seizure class, the distribution of seizure examples and their patient-wise medoids indicates that neither representativeness nor similarity alone is sufficient for optimal selection. In contrast, when the test query corresponds to non-seizure activity, representative non-seizure examples effectively capture the class distribution, whereas representative seizure medoids avoid introducing misleading outliers. These observations demonstrate the necessity of jointly considering both representativeness and similarity in the proposed RAICL framework.

\begin{figure*}[htpb] \centering
\subfigure[]{\includegraphics[width=.35\linewidth,clip]{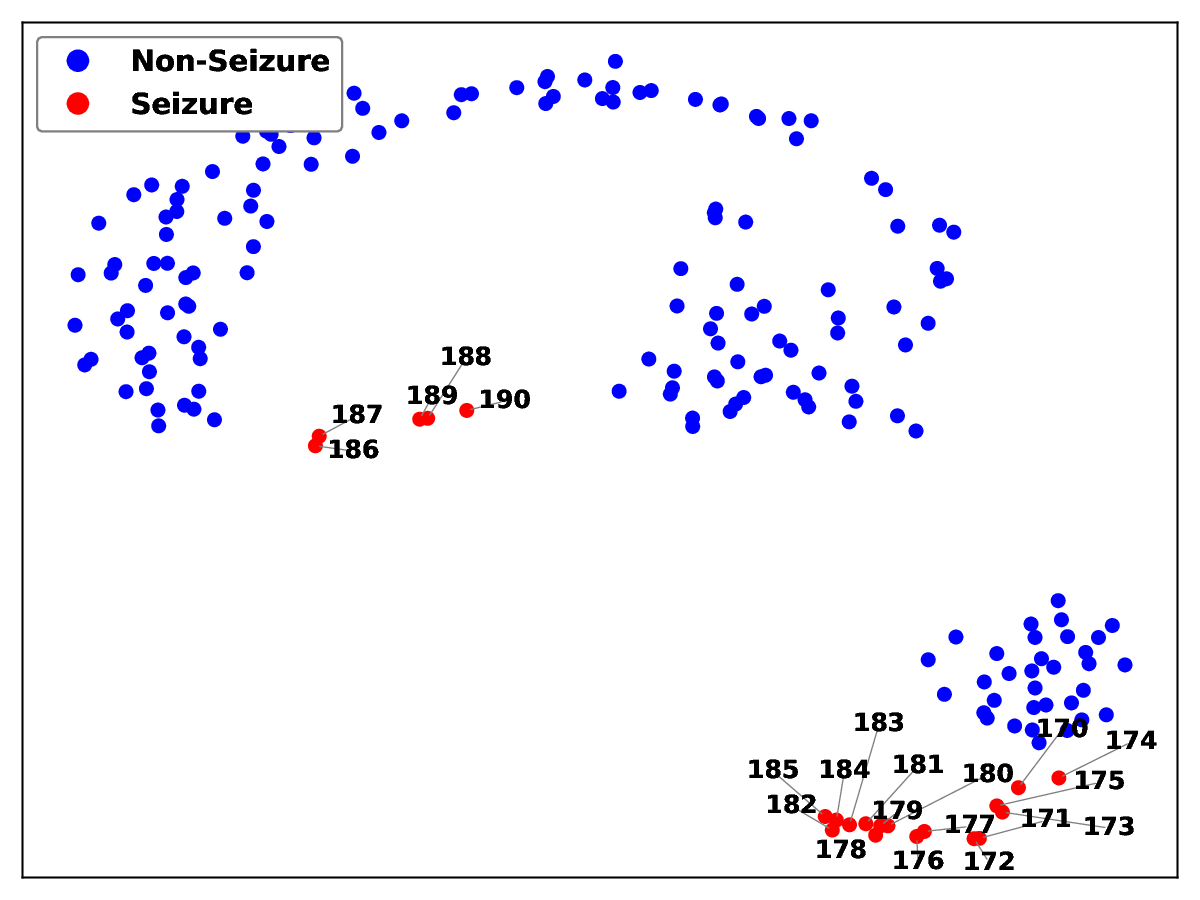}\label{fig:tsne-chromatic}}
\subfigure[]{\includegraphics[width=.35\linewidth,clip]{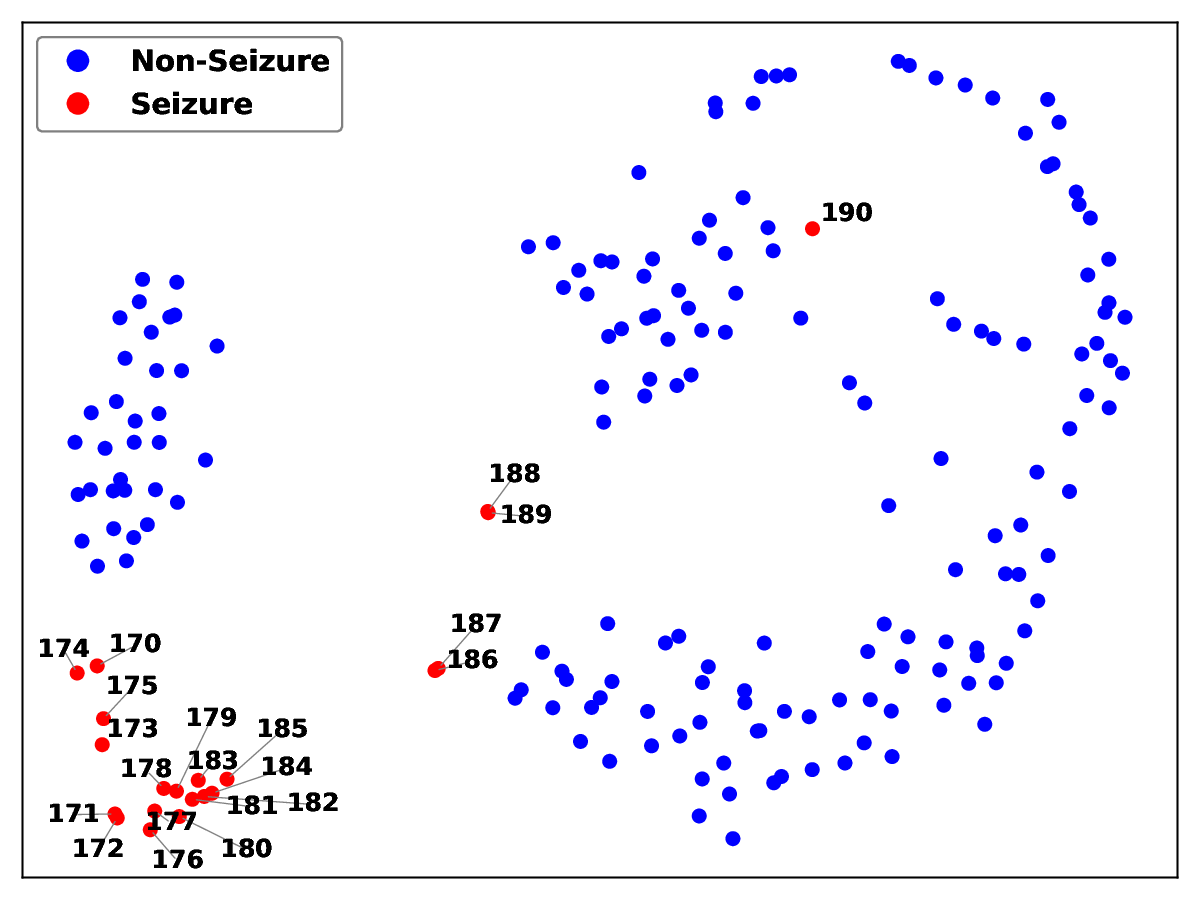}\label{fig:tsne-blackwhite}}
\subfigure[]{\includegraphics[width=.27\linewidth,clip]{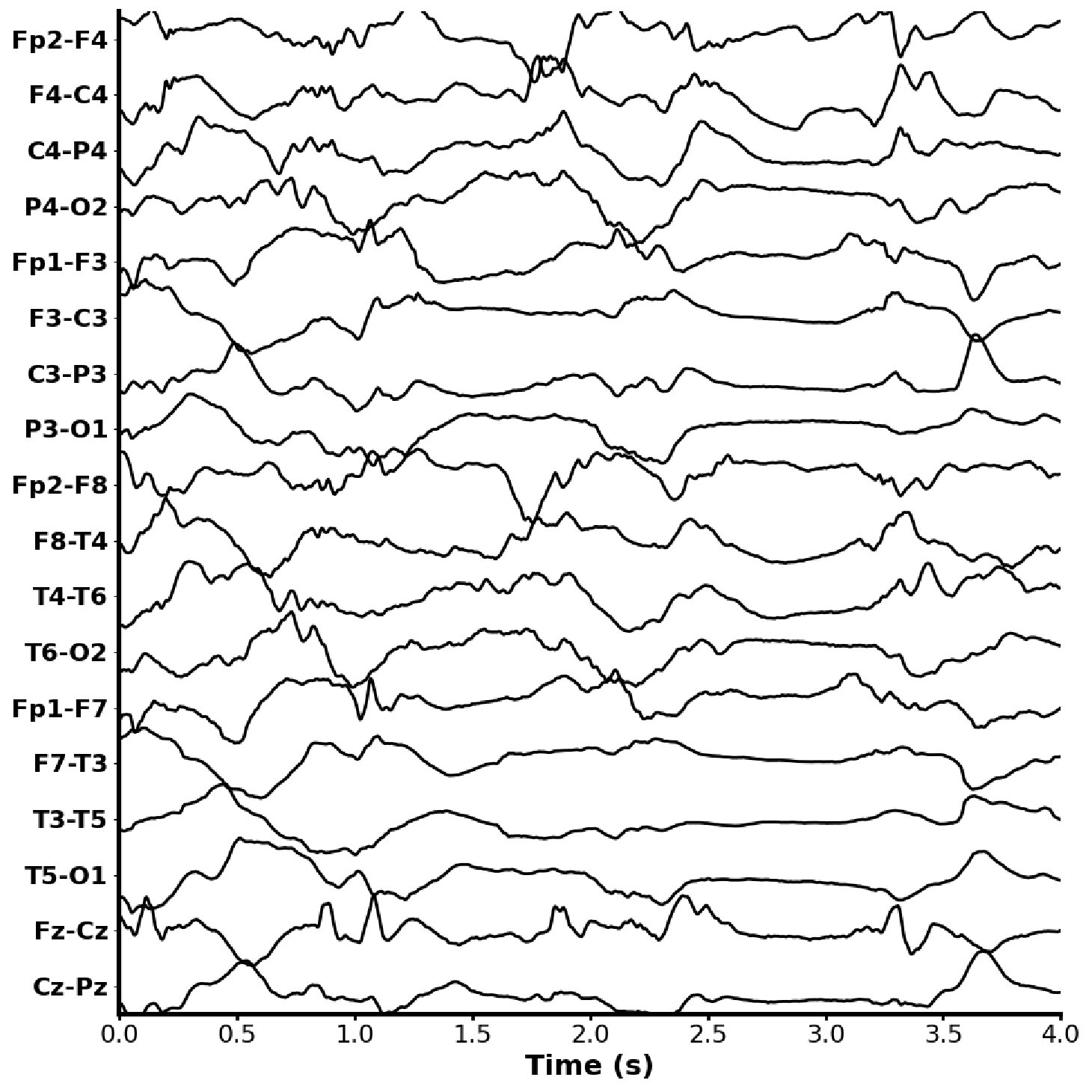}\label{fig:blackwhitetrial}}
\caption{$t$-SNE visualization of VLM input embeddings derived from EEG waveform images with chromatic and monochromatic channel representations. Embeddings are extracted from trials of the first patient in the CHSZ dataset using the Qwen3-VL-32B-Instruct VLM with a reasoning prompt. (a) Chromatic channel representation; (b) Monochromatic (black) channel representation; and (c) The 190th trial plotted with monochromatic channels, identified as an outlier in (b).} \label{fig:chromaticplot}
\end{figure*}

\subsection{Verification on EEG Waveform Plotting} \label{sect:waveformplot}

This subsection evaluates the effectiveness of the three key design choices for EEG waveform plotting introduced in Section~\ref{sect:eegwaveformplot}.

Fig.~\ref{fig:vlm4eeg} illustrates an example response from the Gemini-3-Flash VLM to a seizure query from the NICU dataset. The model correctly interprets electrode annotations embedded in the stacked EEG waveform image and accurately associates them with the corresponding signal traces. Moreover, the model successfully follows the provided textual instructions, indicating effective multimodal comprehension.

To compare chromatic and monochromatic channel plotting, we extract the projected visual embeddings that serve as inputs to the LLM module of the Qwen3-VL-32B-Instruct language model. These embeddings are produced by the projection adapter, which maps visual encoder outputs into the native embedding space of the language model. This alignment enables the LLM to interpret visual patterns as semantic inputs. Consequently, these embeddings encode the mapping of each waveform to its specific electrode identity.

Fig.~\ref{fig:chromaticplot} visualizes the resulting class distributions. A clear separation is observed in Fig.~\ref{fig:tsne-chromatic}, whereas Fig.~\ref{fig:tsne-blackwhite} exhibits a outlier of seizure class that was substantially less distinguishable from non-seizure samples. Fig.~\ref{fig:blackwhitetrial} further illustrates one such outlier trial, characterized by extreme signal amplitudes. Overlapping waveforms across electrode channels obscure electrode attribution, thereby degrading model comprehension of EEG patterns of the embeddings and leading to inferior performance, which proves the superiority of chromatic encoding of channels.

\section{Conclusions}\label{sect:conclusions}
This study presents a multimodal framework for EEG-based seizure detection using large-scale VLMs. It showcases an effective method of converting EEG to stacked waveform plot, and selecting the most important examples for RAICL. While this research primarily focuses on binary classification, the methodology is inherently extensible to more complex tasks. As VLMs continue to evolve with expanded context windows, the integration of a larger number of in-context demonstrations within the prompt range is expected to further enhance decoding performance. Furthermore, the proposed framework is fully compatible with multi-class classification tasks, such as seizure subtype identification~\cite{Peng2022}.

Our proposed paradigm suggests an important paradigm shift from signal-based analysis that is limited to small scale of data. Our experimental results indicate that for EEG waveform decoding, the performance of the VLM is mostly driven by the visual encoder rather than the language model component. These findings suggest that future research should prioritize the development of task-specific visual encoders. Aligning such fine-tuned visual encoders with lightweight language models may provide a more efficient and computationally feasible solution for physiological signal decoding in clinical environments.

\bibliographystyle{IEEEtran} \bibliography{vlm4eeg}

\end{document}